# Electric control of spin injection into a ferroelectric semiconductor

Xiaohui Liu, J. D. Burton[*] and Evgeny Y. Tsymbal[†]

*Department of Physics and Astronomy & Nebraska Center for Materials and Nanoscience,
University of Nebraska, Lincoln, Nebraska 68588-0299, USA*

Electric-field control of spin-dependent properties has become one of the most attractive phenomena in modern materials research due the promise of new device functionalities. One of the paradigms in this approach is to electrically toggle the spin polarization of carriers injected into a semiconductor using ferroelectric polarization as a control parameter. Using first-principles density functional calculations, we explore the effect of ferroelectric polarization of electron-doped $BaTiO_3$ ($n$-$BaTiO_3$) on the spin-polarized transmission across the $SrRuO_3$/$n$-$BaTiO_3$ (001) interface. Our study reveals that the interface transmission is negatively spin-polarized and that ferroelectric polarization reversal leads to a change in the transport spin polarization from -65% to -98%. We show that this effect stems from the large difference in Fermi wave vectors between up- and down-spins in ferromagnetic $SrRuO_3$ and a change in the transport regime driven by ferroelectric polarization switching. The predicted sizeable change in the spin polarization provides a non-volatile mechanism to electrically control spin injection in semiconductor-based spintronics devices.

Spin injection is one of the key phenomena exploiting the electron spin degree of freedom in future electronic devices.[1] A critical parameter that determines the efficiency of spin-injection is the degree of spin polarization which is carried by spin-polarized current. An efficient spin injection into metals has been commercially employed in today's magnetic read heads and magnetic random access memories through the tunneling magnetoresistance effect in magnetic tunnel junctions. Significant interest has been addressed to the spin injection into semiconductors,[2-7] and recent developments in the field have demonstrated a possibility of efficient spin-injection and spin-detection in various electronic systems.[8,9] All the above results rely however on a "passive" spin injection where the degree of transport spin polarization is determined by the spin polarization of the injector and the detector, and the electronic properties of the interface. Adjustable spin injection with a controllable degree of spin polarization would be appealing from the scientific point of view and useful for applications in future spintronic devices.

Recently, experiment and theory have found that ferroelectric polarization can be used to control magnetization at all-oxide ferroelectric/ferromagnetic interfaces.[10-12] Studies in such oxide systems reveal that proper engineering of the interface plays a crucial rule in the manifestation of such novel phenomena.[13] Reversal of ferroelectric polarization provides a bistable mechanism to electrically control electronic systems and this characteristic can be used to design novel electronic devices. Efforts have been made in this field, and an important route taken is where ferroelectric materials are introduced as functional barriers in tunnel junctions,[14,15] providing a possibility to strongly affect the resistance of such a ferroelectric tunnel junction (FTJ) by ferroelectric polarization switching. This functionality of FTJs was predicted to be extended by employing ferromagnetic electrodes[16-18] which stimulated experimental studies and led to a number of demonstrations of tunable spin-polarization of the tunneling current.[19-22]

While ferroelectric materials used in FTJs are normally considered as insulators, previous studies have found that ferroelectricity persists even in moderately electron-doped (i.e. metallic, or nearly so) $BaTiO_3$.[23,24] These results were corroborated by theoretical studies showing that that ferroelectric displacements in $BaTiO_3$ persist up to the doping level of about $0.1e$ per unit cell (~$10^{21}/cm^3$).[25,26] The combination of ferroelectricity and conductivity in one material introduces unique electric properties, opening the door to extended functionalities. In our previous work,[27] we showed that the ferroelectric polarization can be used to alter the resistive nature of the interface between $n$-$BaTiO_3$ and metallic $SrRuO_3$. Specifically, we found that polarization switching in $n$-$BaTiO_3$ induces a transition between Ohmic and Schottky regimes, leading to a five-orders-of-magnitude change in interface resistance.

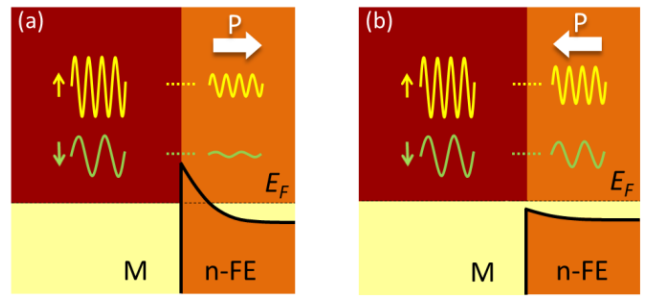

FIG. 1. Polarization controlled band alignment at in the interface between a ferromagnetic metal (FM), e.g. $SrRuO_3$, and electron-doped ferroelectric ($n$-FE), e.g. $n$-$BaTiO_3$. Horizontal arrows indicate the polarization direction. Schottky (a) and Ohmic (b) contacts are created for polarization pointing away from and into the interface respectively. Waves depict incident and transmitted Bloch states for spin-up and spin-down electrons.

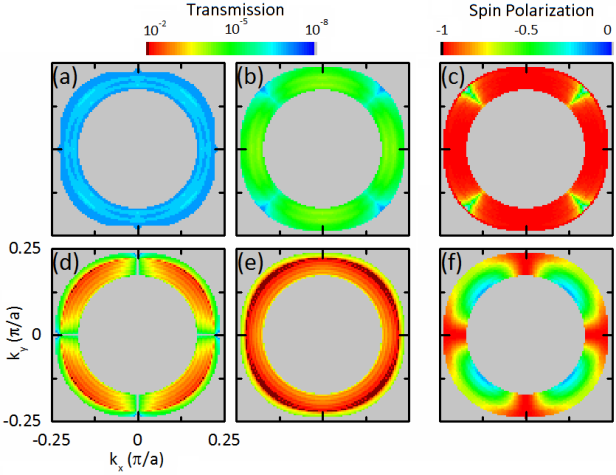

FIG. 2. $\mathbf{k}_{\parallel}$-resolved transmission through the Schottky interface for (a) spin-up and (b) spin-down electrons. (c) $\mathbf{k}_{\parallel}$-resolved spin-polarization for the Schottky interface. Note that transmission is only plotted in a small region around $\mathbf{k}_{\parallel} = 0$, all other points in the 2DBZ have zero transmission. (d-f) Same as in (a-c) for the Ohmic interface.

In this paper we demonstrate that ferroelectric polarization can be used as a control parameter to tune the spin polarization of injected carries from a ferromagnetic (FM) metal to an electron-doped ferroelectric (*n*-FE). As a model system we use a $SrRuO_3/n$-$BaTiO_3$ (001) junction, where we take into account the spin-polarized electronic band structure of $SrRuO_3$. Since $SrRuO_3$ is ferromagnetic below the Curie temperature of 160K,[28] the transmission across such an interface is spin-polarized and the magnitude of this spin-polarization is expected to depend on the orientation of the ferroelectric polarization, as is indicated schematically in Fig. 1. Our calculations confirm this expectation, predicting a significant change in the transport spin-polarization, which is the central result of this work.

First-principles calculations are performed on a supercell as described in our previous work[27] using the plane-wave pseudopotential code QUANTUM ESPRESSO,[29] where the exchange and correlation effects are treated within the local spin-density approximation. We assume that the electron doping of *n*-$BaTiO_3$ is 0.06 *e*/formula unit, which is realized by the virtual crystal approximation[30] applied to the oxygen potentials in $BaTiO_3$. Self-consistent spin-polarized calculations are performed to relax the electronic structure with no additional relaxation of the atomic structure resulting from the non-spin-polarized calculation. Transport properties, i.e. the spin-dependent interface transmission, are calculated using a general scattering formalism implemented in the QUANTUM ESPRESSO.

Consistent with our previous work,[27] we find that reversal of ferroelectric polarization of *n*-$BaTiO_3$ results in a transition between two contact regimes: Schottky and Ohmic. We find, however, that taking into account a spin-polarized band structure of $SrRuO_3$ leads to a smaller change in the interface resistance with polarization reversal, as compared to non-spin-polarized calculations. Specifically, we obtain the total resistance of $0.28\times10^2$ $\Omega\mu m^2$ for the Ohmic contact and $0.95\times10^4$ $\Omega\mu m^2$ for Schottky contact, revealing about three-orders-of-magnitude change in the interface resistance. This difference between spin-polarized and non-spin-polarized results is due to the opening of the spin-down transmission channel in $SrRuO_3$ for the former, which has a larger wave vector and therefore higher probability of tunneling across the Schottky barrier.

For each contact, we calculate transmission for spin-up and spin-down electrons ($T_\uparrow$ and $T_\downarrow$, respectively) over the two-dimensional Brillouin zone (2DBZ), as shown in Fig. 2. The transmission is distributed in a ring-shaped area centered around the $\bar{\Gamma}$ point (i.e. $\mathbf{k}_{\parallel} = 0$). Regions of the 2DBZ with non-zero transmission occur only where the Fermi surface projections of $SrRuO_3$ and *n*-$BaTiO_3$ overlap, leading to the ring-like distribution. For both polarization orientations (i.e. for both interface contact regimes), the spin-down transmission is larger than that of the spin-up transmission. Figs. 2(c) and (f) show the spin-polarization of the interface transmission, which is defined by SP = $(T_\uparrow - T_\downarrow)/(T_\uparrow + T_\downarrow)$ and calculated over the 2DBZ. It is evident that for both contact regimes, the net spin polarization is negative. When ferroelectric polarization is pointing toward the interface and the contact is Ohmic, the net spin polarization is -65%, Fig. 2(f). When the ferroelectric polarization is switched to point away from the interface and the contact is Schottky the spin-polarization in this case is negatively enhanced to -98%, Fig. 2(c).

To understand such a strong effect, we start from examining the Fermi surface of $SrRuO_3$ (Fig. 3). It covers nearly the entire 2DBZ, as seen from Figs. 3(a, b) and 3(c, d) for spin-up and spin-down respectively. The Fermi surface of *n*-$BaTiO_3$ consists of a single sheet forming a corrugated tube oriented along the electric polarization, as shown previously in Ref. 27. The overlap between the Fermi surfaces of $SrRuO_3$ and *n*-$BaTiO_3$, viewed along the transport direction, leads to the ring-like area approximately indicated by the concentric circles in Fig. 3(b) and (d). Since we consider complete in-plane periodicity there is no mixing between different $\mathbf{k}_{\parallel}$ and, therefore, to study the spin-polarized transmission, we need only to take into account the properties of states located in this region of the Fermi surface of $SrRuO_3$. The orbital analysis of these states on the Fermi surface reveals that spin-up states are composed mainly of the $d_{z^2}$ orbital (the yellow surface in Figs. 3(a) and 3(b)), while the spin-down states are composed of $d_{zx}$ and $d_{zy}$ orbitals (the magenta surface in Figs. 3(c) and 3(d)).

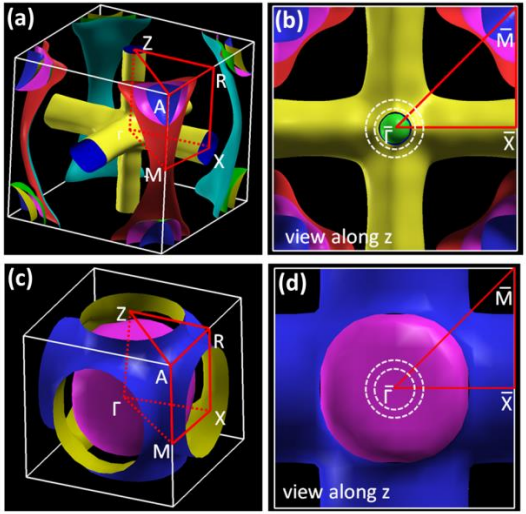

FIG. 3. Fermi surface of spin-up (a) and spin-down (c) and their view along $z$ direction respectively (b) and (d). The concentric rings in (b) and (d) approximately demark the minimum and maximum radius of the Fermi surface of $n$-BaTiO$_3$.

The negative value of spin polarization, as found for both cases, as well as the change in spin-polarization magnitude can be understood using the arguments put forth by Slonczewski.[31] According to the Slonczewski model, first, the spin-polarization of the transmission coefficient for a given $\mathbf{k}_\parallel$ is negative if $k_{z\downarrow}/k_{z\uparrow} > 1$. Second, the magnitude of the spin-polarization depends on the effective barrier height for each $\mathbf{k}_\parallel$: higher barriers lead to an enhanced spin-filtering.

The results of our calculations conform to both of these relationships. The spin-resolved Fermi surfaces of SrRuO$_3$ have quite different characteristics in the ring-like region of the 2DBZ, with $k_{z\downarrow}/k_{z\uparrow} \gg 1$, as seen by comparing the yellow surface for spin-up in Fig. 3(a,b) with the magenta surface for spin-down in Fig. 3(c,d). This behavior can be understood in terms of the orbital character of the spin-dependent states comprising the Fermi surface. The crystal field lowers the energy of the $t_{2g}$ orbitals with respect to the $e_g$ orbitals. This reduces the potential energy of the spin-down $d_{zx}$ and $d_{zy}$ states and, hence, enhances their kinetic energy on the Fermi surface, which is reflected in a nearly spherical Fermi surface and a larger Fermi wave vector for the spin-down states. On the contrary, the higher energy of the spin-up $d_{z^2}$ states strongly affects the shape of the Fermi surface causing it to form a cross pattern of three corrugated tubes, leading to small values of the Fermi wave vector in the vicinity to the $\Gamma$ point for the spin-up states.

When the ferroelectric polarization of the $n$-BaTiO$_3$ points into SrRuO$_3$, as shown in Fig. 1(b), the Fermi level is located closer to the bottom of conduction bands of $n$-BaTiO$_3$ than in the bulk. This leads to the first layer of $n$-BaTiO$_3$ near the interface being, in fact, an effective tunneling barrier, despite the small occupation of the conduction band. When ferroelectric polarization is reversed to point away from SrRuO$_3$, as shown in Fig. 1(a), there is complete depletion of conduction band states near the interface (i.e. a Schottky barrier) and hence the height of the tunneling barrier is dramatically increased.

We conclude therefore that the negative spin-polarization can be explained by the existence of a tunneling barrier at the SrRuO$_3$/$n$-BaTiO$_3$ interface and the spin-dependent Fermi surface of SrRuO$_3$ which is characterized by a larger wave vector for spin-down electrons compared to spin-up electrons ($k_{z\downarrow}/k_{z\uparrow} > 1$). Furthermore, when the ferroelectric polarization is reversed from pointing into the interface to pointing away from the interface the dramatic increase in the barrier height leads to the substantial enhancement in the magnitude of the spin-polarization, consistent with the Slonczewski model.

The predicted change in the transport spin-polarization with polarization reversal is also reflected by the induced local density of states within the $n$-BaTiO$_3$ barrier near the interface. Fig. 4 shows the spin-polarized local density of states on the interfacial Ti atom for both contact regimes. It is seen that, within the transmission ring, the induced density of states is more negatively spin-polarized for the Schottky contact than for the Ohmic contact. This observation is consistent with our prediction of the enhanced negative spin polarization in the Schottky contact regime.

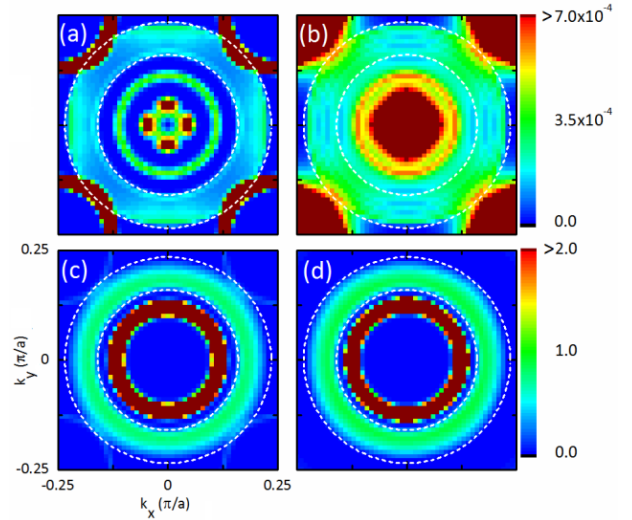

FIG. 4. Spin-up (a, c) and spin-down (b, d) $\mathbf{k}_\parallel$-resolved local density of states on the interfacial Ti atom for Schottky (a, b) and Ohmic (c, d) contacts.

The predicted ferroelectrically-tunable transport spin polarization is not limited to the particular SrRuO$_3$/$n$-BaTiO$_3$ junction considered in this work. We expect the phenomenon to be a general feature of the FM/$n$-FE interface. Moreover, we anticipate the possibility of spin-polarization control over a broader range of values, including a change between positive and negative. This additional tunability can be achieved by changing the doping level on the ferroelectric, as well as using interface engineering to adjust the Schottky barrier at the interface [32,33] and/or enhance ferroelectric polarization stability.[34] The detection of spin polarization may be achieved using methods similar to those adopted in the studies of spin injection into semiconductors.[2-7]

In summary, we have shown that a ferromagnet/$n$-doped ferroelectric junction can be used to control the spin-polarization of injected carries. For the prototypical SrRuO$_3$/$n$-BaTiO$_3$ junction, we predicted that reversal of ferroelectric polarization of $n$-BaTiO$_3$ changes the spin-polarization of transmission from -65% to -98%. This sizable change occurs due to the effect of ferroelectric polarization on the effective contact barrier height which selects preferentially electrons with a certain spin orientation as a result of the spin-dependent Fermi surface of SrRuO$_3$. The proposed ferroelectrically-tunable spin-polarization offers an exciting prospect to extend the functionalities of semiconductor-based spintronic devices.

This research was supported U.S. Department of Energy, Office of Basic Energy Sciences, Division of Materials Sciences and Engineering (DOE Grant DE-SC0004876). Computations were performed at the University of Nebraska Holland Computing Center.

*E-mail: jdburton1@gmail.com
†E-mail: tsymbal@unl.edu